# Origin of interface magnetism in BiMnO$_3$/SrTiO$_3$ and LaAlO$_3$/SrTiO$_3$ heterostructures

M. Salluzzo,[1, *] S. Gariglio,[2] D. Stornaiuolo,[2] V. Sessi,[3] S. Rusponi,[4] C. Piamonteze,[5] G. M. De Luca,[1] M. Minola,[6] D. Marré,[7] A. Gadaleta,[7] H. Brune,[4] F. Nolting,[5] N. B. Brookes,[3] and G. Ghiringhelli[6, †]

[1]CNR-SPIN, Complesso MonteSantangelo via Cinthia, I-80126 Napoli, Italy
[2]Département de Physique de la Matière Condensée, University of Geneva,
24 Quai Ernest-Ansermet, CH-1211 Geneva 4, Switzerland
[3]European Synchrotron Radiation Facility, 6 Rue Jules Horowitz, B.P. 220, F-38043 Grenoble Cedex, France
[4]Institute of Condensed Matter Physics, Ecole Polytechnique Fédérale de Lausanne, CH-1015 Lausanne, Switzerland
[5]Swiss Light Source, Paul Scherrer Institut, CH-5232 Villigen PSI, Switzerland
[6]CNR-SPIN and Dipartimento di Fisica, Politecnico de Milano,
Piazza Leonardo da Vinci 32, I-20133 Milano, Italy
[7]CNR-SPIN and Dipartimento di Fisica, Università di Genova, Via Dodecaneso 33, I-14146 Genova, Italy

Possible ferromagnetism induced in otherwise non-magnetic materials has been motivating intense research in complex oxide heterostructures. Here we show that a confined magnetism is realized at the interface between SrTiO$_3$ and two insulating polar oxides, BiMnO$_3$ and LaAlO$_3$. By using polarization dependent x-ray absorption spectroscopy, we find that in both cases the magnetism can be stabilized by a negative exchange interaction between the electrons transferred to the interface and local magnetic moments. These local magnetic moments are associated to magnetic Ti$^{3+}$ ions at the interface itself for LaAlO$_3$/SrTiO$_3$ and to Mn$^{3+}$ ions in the overlayer for BiMnO$_3$/SrTiO$_3$. In LaAlO$_3$/SrTiO$_3$ the induced magnetism is quenched by annealing in oxygen, suggesting a decisive role of oxygen vacancies in this phenomenon.

The progress in oxide thin film technology is opening the possibility of electronic applications based on the peculiar physical properties of oxide interfaces [1, 2]. The best known example is the junction between two non-magnetic band insulators, LaAlO$_3$ (LAO) and SrTiO$_3$ (STO), which hosts a quasi-two-dimensional electron gas (q2DEG) [3]. The functional properties of LAO/STO heterostructures are indeed extraordinary, such as the possibility of driving an insulating to metal transition by electric field gating at room temperature [4]. At the same time, the real ground state properties of this system are still hotly debated due to apparently conflicting observations of ferromagnetic ordering in some samples [5], and of superconductivity of the q2DEG below 0.3 K [6] in other samples. More recent studies [7–10] have suggested the coexistence of magnetism and superconductivity, quite intriguing for fundamental physics. However, the mere existence and the real nature of this magnetism in LAO/STO are still questioned and urge clarification.

According to some theoretical studies [11, 12], n-type titanate heterostructures can become metallic and ferromagnetic via an electronic reconstruction transferring electrons to the otherwise empty titanium 3d-sites at the interface. Electron-electron correlations usually favor antiferromagnetic-insulating ground states, as in LaTiO$_3$, a prototype Mott insulator with electrons localized at Ti$^{3+}$ sites with 3d$^1$ configuration. However, ferromagnetism can be artificially stabilized at the interfaces as a result of the breaking of symmetry, which lifts the 3d orbital degeneracy. This reconstruction takes place in LAO/STO heterostructures [13], where the electrons transferred to the interface occupy preferentially 3d$_{xy}$-derived bands. Torque [9] and SQUID (Superconducting Quantum Interference Device) [7] magnetometry, and scanning SQUID microscopy [10] experiments have reported evidence of a magnetic state developing at low temperatures. However, these techniques cannot provide the direct proof that magnetism is indeed an intrinsic phenomenon related to Ti$^{3+}$ moments at the interface. Moreover, recent theoretical reports suggested a negligible spin polarization for ideal LAO/STO [14], while magnetism could be induced by the addition of point defects [15], and in particular by oxygen vacancies [14]. On the other hand, an interface magnetic state can be also achieved by exploiting the magnetic proximity effect, i.e. by replacing LaAlO$_3$ with a ferromagnetic polar insulating film, like BiMnO$_3$.

We have studied the nature and the mechanism of the induced magnetism at the interface between STO and polar-overlayers, namely in BiMnO$_3$/SrTiO$_3$ and LaAlO$_3$/SrTiO$_3$ systems. We have used polarization dependent x-ray absorption spectroscopy (XAS) across the transition metals L$_{2,3}$ edges to probe directly the magnetic and orbital properties of Ti at the interface. XAS performed with circularly or linearly polarized photons can detect the magnetic moments and the 3d orbital energy splitting, respectively (see Fig. 1 for a sketch of the experimental setup). The two techniques, usually known as x-ray magnetic circular dichroism (XMCD) and x-ray linear dichroism (XLD) are so sensitive that they can be used on single interfaces [13] and ultra-diluted magnetic impurities [17]. The results presented below give evidence that magnetism at the interface is driven by the exchange interaction between interface states and localized moments either provided by Mn$^{3+}$ across the interface or by oxygen-vacancy induced Ti$^{3+}$ ions at the interface itself.

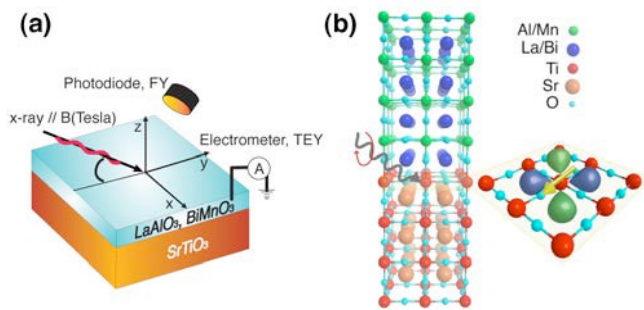

FIG. 1: (color online) Linear and circular dichroism in the XAS of SrTiO$_3$ interfaces. (a) Schematics of the experimental set-up: by absorption of a photon (zig-zag red arrow) of appropriate energy and known polarization, a Ti or Mn 2p electron is promoted to the 3d states. The external magnetic field, **B**, is always parallel to the beam direction and the sample can be oriented at normal (0 degrees) or 70 deg incidence. (b) The crystal structure of an ideal 4 unit cell polar insulating oxide film (LaAlO$_3$ or BiMnO$_3$) deposited on TiO$_2$ terminated SrTiO$_3$ single crystal, and a pictorial view of the outcomes of Ti L$_{2,3}$ XMCD and XLD, which provide insight on the symmetry and occupation of lowest-lying 3d states (Ti-3d$_{xy}$ orbitals in the picture), and on the consequent magnetic moments (yellow arrow).

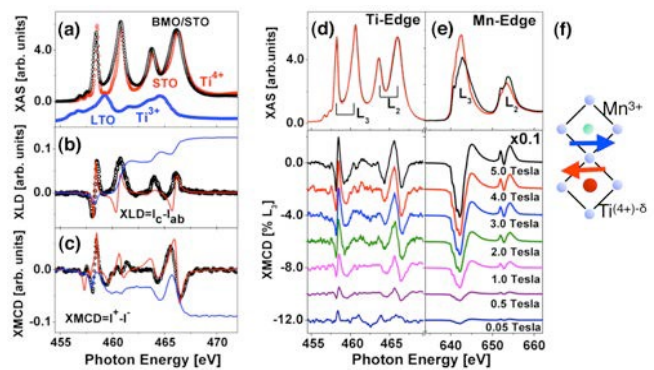

FIG. 2: (color online) (a) Ti XAS spectra of BMO(4nm)/STO (black circles), undoped STO crystal (red circles) and bulk LaTiO$_3$ (blue circles); (b) XLD spectra at grazing incidence (black circles) and integrated XLD (blue line) of BMO/STO, compared to the spectrum calculated in single ion crystal field model (red line) (splitting (t$_{2g}$)=-50 meV and (e$_g$)=-100 meV); (c) same for XMCD at 4 T, where the theoretical spectra include also charge transfer and a magnetic exchange energy (-10 meV, see text). (d,e) Ti and Mn XMCD spectra of BMO/STO as function of magnetic field (bottom panels, offset for clarity) and the typical XAS spectra at 5 T and 8 K (upper panels, red and black curves). (f) Schematics of the charge-transfer process from Mn$^{3+}$ to a Ti$^{4+}$ and arrangement of their magnetic moments at the interface.

LAO(10-unit cells)/STO heterostructures were prepared by Pulsed Laser Deposition (PLD) assisted by RHEED (Reflective High Energy Electron Diffraction) at 800°C in oxygen atmosphere ($1 \times 10^{-4}$ mbar) as in Ref. 18. With the purpose of clarifying the role of oxygen vacancies, after growth some samples were cooled down in the same O$_2$-pressure used during the deposition, while others were in-situ post-annealed for 60 minutes at 530°C in 200 mbar of oxygen in order to eliminate oxygen vacancies naturally present in the non-annealed systems grown by PLD. Epitaxial BMO/STO films were prepared by radio-frequency magnetron sputtering in a total pressure of 0.20 mbar (a mixture of argon and oxygen with a 10:1 ratio), and at temperatures of 650°C. All the samples were fully characterized by x-ray diffraction, electrical transport and atomic force microscopy, some of these results are shown in the on-line supporting materials [19]. The Ti L$_{2,3}$ XMCD signal was obtained as difference between the average of 32 XAS spectra acquired with magnetic field parallel and antiparallel to the photon-helicity vector orientations. The 64 XAS-data needed for each XMCD were collected in a sequence alternating reversal of field and polarization at each spectrum. This procedure ensure the best cancellation of spurious effects. The experiments were performed and validated at two different facilities, namely at the beamlines ID08 of the European Synchrotron Radiation Facility (ESRF) and X-Treme [16] of the Swiss Light Source (SLS).

In Fig. 2 we show XMCD and XLD measured in grazing incidence conditions (70 degrees from the surface normal) on a BMO(10-unit cells)/STO single interface. BiMnO$_3$ is a ferromagnetic (T$_c$ =100 K) Mott-Hubbard insulator [20]; STO is a band insulator, characterized by a 3.2 eV gap between the valence band, with a predominant O2p character, and the conduction band, mainly formed by Ti3d states. As titanium is formally tetravalent (3d$^0$) no magnetism on the Ti atoms is expected for this system. Moreover there is no evidence of a q2DEG at the interface, despite the polar structure of BMO. Nevertheless, in Fig. 2a we show that electrons have been actually transferred to the STO interface layers. Indeed, the comparison of XAS spectrum of BMO/STO with that of bare STO [21] reveals an extra intensity in the valleys between the main Ti$^{4+}$ peaks (around 459.5 eV and 462 eV), which means that the effective Ti valence is substantially lower than 4 in agreement with similar manganite/titanate interfaces [24]. By looking at the XLD spectra (Fig. 2b) we can see that the lowest lying 3d states of Ti have in-plane orientation, similarly to what had been found for the LAO/STO interface [13]. Moreover the integral of XLD has positive sign, corresponding to a preferred occupation of 3d$_{xy}$ orbitals.

The described transfer of electrons to the interfacial STO layers is accompanied by a spin splitting of the Ti3d states, as demonstrated by the non-zero XMCD spectra of Fig. 2c. Remarkably, this magnetic signal is genuinely originated at the interface, as demonstrated by the fact that XMCD, sizable when measured by surface sensitive total electron yield as shown in Fig. 2, falls below the detection limit with bulk sensitive fluorescence yield detection [19]. Although the spin moment sum rule fails for Ti, the one on orbital moment can be safely used [22]. From the integral of the total electron yield XMCD (blue line in Fig. 2c) we obtain a net orbital magnetic moment of $m_{orb}$=0.035 $\mu_B$/Ti parallel to the applied magnetic field, as averaged over the probing depth (4 nm) [19]. According to Hund's rules this implies that the spin moment is anti-parallel to the magnetic field and to the Mn spin moment as obtained from XMCD of Fig. 2e. The finite atomic magnetic moment confirms that Ti at the interface has non-zero 3d population. One might then expect that the XMCD spectrum would be given by the "magnetic" $3d^1$ configuration only, but this is not true in the measured spectra. The XMCD signal of Fig. 2c-d is dominated by features at energies of the $3d^0$ peaks in the XAS spectrum. This $3d^0$-like contribution is possible because electrons excited in the $2p^5 3d^1$ XAS final states are sensitive to a magnetic environment through an exchange field. To confirm this interpretation we have simulated the spectra with a multiplet splitting atomic model ($Ti^{4+}$ - $Ti^{3+}$ configuration interaction), including octahedral crystal field and O2p →Ti3d charge-transfer ($\Delta_{CT}$ = 2 eV, equivalent to an effective titanium valence of 3.6±0.05 [19]). Theoretical XAS spectra are dominated by $Ti^{4+}$ peaks and their XLD and XMCD spectra are in good agreement with experimental ones (Fig. 2b,c, thin red line). Interestingly, the sign of dichroism corresponds to a negative exchange interaction, confirming the antiparallel orientation of Ti and Mn magnetic moments (Fig. 2f). We notice also that the amplitude of the Ti XMCD vs. the magnetic field at 8K strictly follows that of Mn. Moreover, the temperature dependence of the XMCD spectra at Ti and Mn follow the same ferromagnetic-like behavior, with $T_C$ close to 100 K (see the supplementary materials [19]). These results confirm that Ti magnetism is associated to the negative Mn-Ti super-exchange interaction, in agreement with previous results on other manganite/titanate multilayers [23]. The fact that the BMO/STO interface is not metallic despite a sizable Ti 3d population is probably due to correlation effects and has some analogies to $Ti^{3+}$ compounds like $YTiO_3$. We can see this in terms of Ti3d-O2p-Mn3d hybrid states at the interface lying in the STO gap, well below the chemical potential level [25]. Therefore in BMO/STO all Ti sites at the interface are magnetic.

Since the interface ferromagnetism found in BMO/STO is due to the coupling between the localized $Mn^{3+}$ spin and the (localized) electrons transferred to the interface states, no magnetism would be expected by replacing $BiMnO_3$ with the non-magnetic $LaAlO_3$. However, also in LAO/STO we have detected a Ti-related magnetic signal dependent on the oxygen stoichiometry, which is, in turn, controlled by the oxygen annealing process. We observe important magnetic and electronic differences between $O_2$-annealed and non-annealed LAO/STO, as shown in Fig. 3. Oxygen vacancies bring a sizable fraction of partially occupied 3d states of Ti sites similarly to the BMO/STO case, as revealed by the XAS spectra of Fig. 3a. On the contrary the annealed interface has a pure $Ti^{4+}$ spectrum, similar to that of bare STO (Fig. 3b). These spectra suggest the amount of interface electrons to be much higher in non-annealed samples [26]. However although the carrier density measured by Hall-effect is indeed lower in annealed LAO/STO than in non-annealed ones (2-4×$10^{13}$ cm$^{-2}$ and 8-10×$10^{13}$ cm$^{-2}$ respectively), the change is insufficient to explain the spectroscopic differences. This means that not all the electrons accumulated at the interface are mobile and contribute to transport. We also find that the 3d orbital splitting has the same sign in the two types of samples, but it is made larger by the oxygen defects, as shown by the XLD of Fig. 3c,d [13]. Finally we can see in Fig. 3e,f and in Fig. 4a that a magnetic signal, much smaller than in BMO/STO, is detected in both cases, but only in non-annealed samples the integral of the XMCD is different from zero. The corresponding orbital moment is smaller by a factor 6-8 to that of BMO/STO, but still quite sizable considering its pure interface character. Moreover, it saturates above 3 T (Fig. 4c), indicating a (weak) ferromagnetic behavior at 8K and below.

The two LAO/STO systems differ also in the shape of the XMCD: the presence of oxygen vacancies in non-annealed LAO/STO brings extra spectral features that we can assign to the $3d^1$ magnetic component. This assignment is confirmed by the atomic multiplet calculations for $3d^1$ ($Ti^{3+}$) and $3d^0$+CT ($Ti^{(4-\delta)+}$) configurations, shown in Fig. 4b. In order to reproduce the correct sign of these features, we had to use a positive exchange interaction for the $3d^1$ configuration and a negative exchange for the $3d^0$-like component. This implies that a non-negligible fraction of the 3d electrons introduced by oxygen vacancies lead to pure $Ti^{3+}$ $3d^1$ local magnetic moments parallel to the external magnetic field, together with a sizable anti-parallel $Ti^{(4-\delta)+}$ component behaving similarly to the $3d^0$-like configuration in BMO/STO. Thus, in LAO/STO the $Ti^{3+}$ magnetic moments play a role similar to $Mn^{3+}$ in BMO/STO, in that they couple antiferromagnetically to the remaining electrons, including those forming the q2DEG.

For LAO/STO annealed in $O_2$ we observe a quenching of the induced XMCD signal, which is characterized by weak features corresponding only to the $Ti^{4+}$ $L_3$ $t_{2g}$ peak (Fig. 3f). Noticeably a signal persists down to fields as low as 0.1 T, indicating that the q2DEG is very slightly spin-polarized [27]. However, the orbital moment is very small ($\ll 0.01$ $\mu_B$/Ti within the probe depth) even at 6 T and 3 K, and also the integral of just the $L_3$ range (452-462 eV) is vanishing. Thus, also the spin moment is very small ($< 10^{-3}$ $\mu_B$/Ti) suggesting that, in agreement with Pavlenko et al. in Ref. 14, the q2DEG is not intrinsically magnetic. The data are compatible with the upper limit given by neutron reflectometry at 11 T and 1.5 K [28], but disagree with that reported recently by Lee et al. for LAO/STO at 0.2 T and 10 K [29]. The average magnetic moment of annealed LAO/STO samples is probably non-zero but it is so small that it can be attributed, in our case, to residual oxygen vacancies or other imperfections, like cationic defects [30], which introduce local $Ti^{3+}$ magnetic moments coupled antiferromagnetically to the q2DEG. Thus, our results are incompatible with the general presence of magnetic $3d^1$ states ($Ti^{3+}$) at the LAO/STO conductive interfaces: the corresponding XMCD shape would have been different from that measured on our samples and its integral would have been non-zero irrespective of the presence of oxygen vacancies.

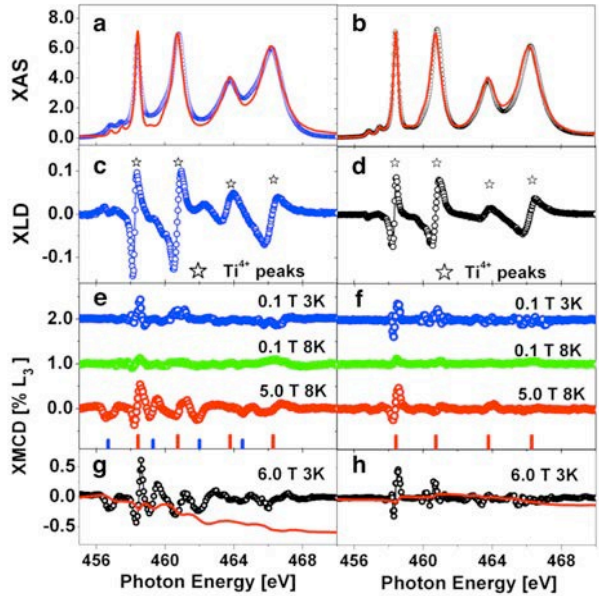

FIG. 3: (color online) X-ray linear and circular dichroism of LAO/STO interfaces. XAS spectra of similar LAO(10-uc)/STO samples, one (a) containing oxygen vacancies (not-annealed, blue circles) and one (b) annealed in oxygen (standard LAO/STO, black circles); insulating STO (red line) is shown as reference. Corresponding XLD (c,d) and XMCD (e,f) spectra. Stars in XLD and red sticks in XMCD indicate the energy of the main $Ti^{4+}$ XAS peaks, while blue sticks indicates those of the $Ti^{3+}$ XAS. The XMCD spectra measured at lowest T and highest B have been integrated leading to a sizable orbital moment in the vacancy rich sample (g) and to an almost vanishing value for the defect free sample (h).

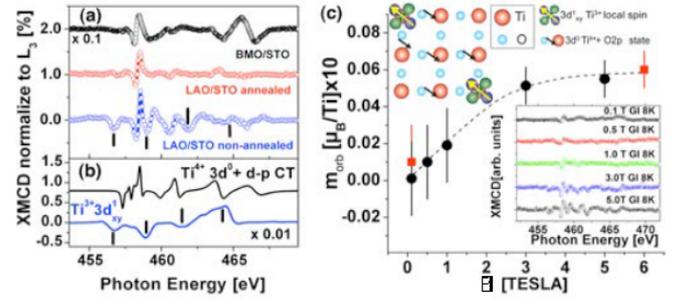

FIG. 4: (color online) Magnetism in LAO/STO heterostructures: (a) Normalized Ti XMCD of BMO/STO (black circles, scaled by a factor 0.1), of standard LAO/STO (red circles) and of oxygen-vacancy rich LAO/STO (blue open circles). (b) Single ion calculations of the XMCD spectra for $Ti^{4+}$ (black line) and $Ti^{3+}$ (blue line), using an exchange field of -10 meV and +10 meV respectively. (c) The field dependence of the orbital moment at 8K (black circles) and at 3K (red squares) as obtained from the XMCD integral for oxygen-vacancy-rich LAO/STO (some of the corresponding spectra in the inset). Cartoon: in LAO/STO oxygen vacancies induce spin and orbital moments at the $Ti^{3+}$ sites; the q2DEG gets polarized by these magnetic impurities and mediates the magnetic interaction.

Titanate heterostructures hold interesting analogies with Mn-doped GaAs dilute magnetic semiconductors (DMS), where Mn ions have both the role of carriers (holes) and magnetic dopants [31]. The ferromagnetism of these materials is explained by a mean-field Zener model, where delocalized holes in hybridized As4p and Mn3d states mediate a long-range ferromagnetism through a negative exchange (p − d exchange) interaction with localized Mn magnetic moments [32]. As sketched in Fig. 4c, in our LAO/STO heterostructures, the $Ti^{3+}$ localized magnetic moments, introduced by oxygen vacancies, play the role of $Mn^{2+}$ impurities in a DMS, while the electrons transferred in mixed O2p-Ti3d states can mediate a magnetic order (Fig. 4c). However, the density of magnetic $Ti^{3+}$, even in the case of non-annealed LAO/STO, appears insufficient to allow a robust long-range ferromagnetism. Yet we believe that the same kind of interface magnetism can settle also in other electronically reconstructed oxide heterostructures, possibly characterized by higher induced magnetic moments.

The authors are grateful to Jean-Marc Triscone, Jochen Mannhart and Mathieu Le Tacon for useful discussions. We also acknowledge P. Zubko for the characterization of $BiMnO_3$/$SrTiO_3$ films. This research was funded by the European Union Seventh Framework Program under Grant Agreement No. 264098-MAMA and from the Italian MIUR Grant No. PRIN 20094W2LAY.

* Electronic address: marco.salluzzo@spin.cnr.it
† Electronic address: giacomo.ghiringhelli@polimi.it